# Topologically protected pseudospins in 2D spring-mass system

Yun Zhou,[1] Prabhakar R. Bandaru,[1,*] and Daniel F. Sievenpiper[2,*]

[1]*Department of Mechanical and Aerospace Engineering, University of California, San Diego, La Jolla, California 92093, USA*

[2]*Department of Electrical and Computer Engineering, University of California, San Diego, La Jolla, California 92093, USA*

It is proposed that a lattice, with constituent masses and spring constants, may be considered as a model system for topological matter. For instance, a relative variation of the inter- and intra-unit cell spring constants can be used to create, tune, and invert band structure. Such an aspect is obtained while preserving time reversal symmetry, and consequently emulates the quantum spin Hall effect. The modal displacement fields of the mass-spring lattice were superposed so to yield pseudospin fields, with positive or negative group velocity. Considering that harmonic oscillators are the basis of classical and quantum excitations over a range of physical systems, the spring-mass system yields further insight into the constituents and possible utility of topological material.

## I. INTRODUCTION

Inspired by the discovery of topological phases and edge states in electronic materials [1,2], the possibility of building related devices for the control of the propagation of light [3–9] and sound [10–18] is being extensively studied. The related device building blocks may harness three major types of topological phases analogous to those in condensed matter systems: quantum Hall effect (QHE) [19,20], quantum spin Hall effect (QSHE) [21–23], and quantum valley Hall effect (QVHE) [24–27]. The QHE has chiral edge modes, and requires an external magnetic field to break time reversal symmetry (TRS), which may be accomplished in acoustic and photonic systems by adding gyroscopic material or external circulators [3,10–12,28]. The QSHE is amenable to TRS, associated with a pair of spin-locked helical modes, and is obtained by introducing strong spin-orbit coupling [5,8,13,17,18]. The QVHE generates valley-locked chiral edge states, and exploits the valley degrees of freedom [6,29].

*Corresponding authors: pbandaru@ucsd.edu, and dsievenpiper@eng.ucsd.edu

It would of much advantage and yield insight, to consider a harmonic oscillator point of view, quite common in physics, for invoking topological phases. In this respect, a discrete spring-mass based mechanical system, may constitute a model system for topological structure as related to phononic materials. For instance, QHE based topological insulators in spring-mass lattices may be created by adding circulating gyroscopes [11,28], Coriolis force [30] or varying spring tension [31]. QVHE has also been realized in such systems by alternating the mass at A and B sites of the unit cell of a mechanical graphene-like lattice [29]: **Figure 1(a)**, while coupled pendula [32], and a mechanical granular graphene system [33] may mimic QSHE-like phenomena.

In this paper, we propose a two-dimensional (2-D) spring-mass system, exemplifying a QSHE topological insulator, in the acoustic domain. Various trivial and non-trivial band structures may be originated by varying the masses ($m$) and the relative spring constants ($k$) in the associated lattice. In addition to exhibiting the topological features that have now become familiar to practitioners in the field, we indicate a novel *spin* degree of freedom. The related pseudospins are observed, in frequency domain analysis as the polarization of modal displacement field of masses in one unit cell: **Fig. 1 (a).** TRS protected edge modes, incorporating the propagation of such pseudospins, are shown to exist. This structure can be applied as one of the possible practical designs of photonic/phononic topological insulators.

A basis for creating a topological material, based on a spring-mass system, to mimic the QSH

effect, is to create intrinsic TRS. We consider a hexagonal lattice of masses and springs arranged in $C_6$ symmetry. The $E$ and $E'$ representations are each two-fold degenerate with the individuals being complex conjugates [34]. Consequently, a four-fold degeneracy is required to satisfy TRS and may be enabled through manifesting a double Dirac cone in the band structure. We achieve a four-fold degeneracy, in the band structure of a spring-mass constituted lattice by the zone-folding method [8].

## II. THE SPRING-MASS SYSTEM MODEL AND COMPUTATIONAL METHODS

We consider a hexagonal lattice with equal masses $m$ connected by linear springs $k$, as shown in **Fig. 1 (a)**. The unit cell of this hexagonal lattice consists of 2 masses $m^1 = m^2 = m$, with lattice constants $\vec{a_1}$ and $\vec{a_2}$ ($|\vec{a_1}| = |\vec{a_2}| = a$). From Newton's law, the governing equation $M\ddot{u} = F(u)$, where $M$ is a diagonal matrix with the values of the two masses on its diagonal: $M = \text{diag}\{m^1, m^1, m^2, m^2\}$. $u$ is a vector constituted from the two degrees of freedom for each mass – the $x$ and $y$ direction displacements for $m^1$ and $m^2$: $u = \{u_x^1, u_y^1, u_x^2, u_y^2\}$ and $F$ is the force. We consider a Bloch wave solution of the type $u = Ue^{i(qa_1\gamma_1 + la_2\gamma_2 - \omega t)}$ to the governing equation of the $(q, l)^{\text{th}}$ unit cell, where $U = \{U_x^1, U_y^1, U_x^2, U_y^2\}$ is the modal displacement, and $\gamma_1$ and $\gamma_2$ are wave vectors. A dispersion relation is obtained by solving the eigenvalue problem $D(\gamma_1, \gamma_2)U = \omega^2 MU$, with $D$ as a dynamical matrix (see Supplementary Material).

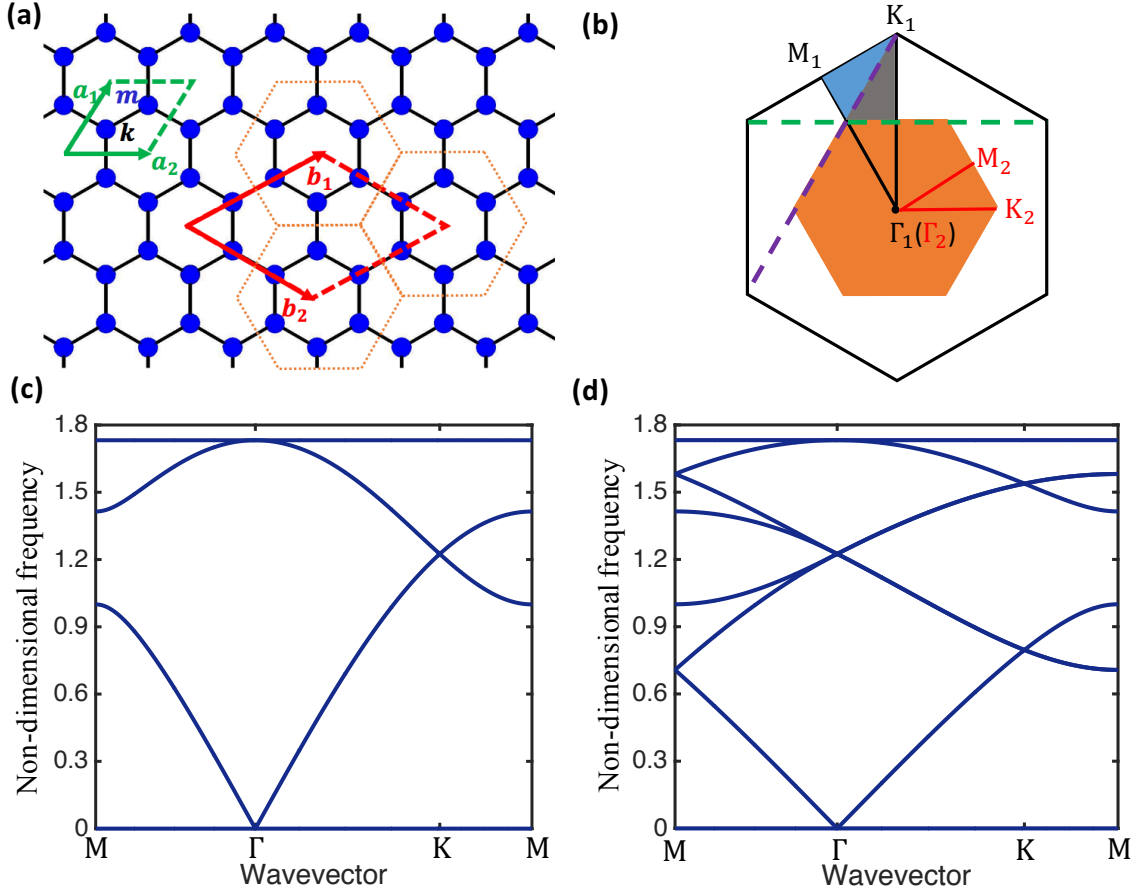

**Figure 1 (a)** Hexagonal spring-mass lattice with uniform spring constant $k$ and mass $m$. $a_1$ and $a_2$ are lattice constants of the unit cell before zone folding, and $b_1$ and $b_2$ are lattice constants of the unit cell after zone folding. **(b)** First Brillouin zone (BZ) before (big hexagon) and after (small hexagon in orange) zone folding. When looking at 1/12 BZ, the triangle $\Gamma_1 M_1 K_1$ is first folded along the purple dashed line, then folded along the green dashed line. **(c)** Band diagram of the lattice in (a) for unit cell of 2 masses, and **(d)** band diagram of the lattice in (a) for the expanded unit cell of 6 masses.

The band structure of the hexagonal lattice in **Fig. 1(c)** exhibits a single Dirac cone at the K (K′) point. The frequencies are non-dimensionalized as $\Omega = \frac{\omega}{\sqrt{\frac{k}{m}}}$. Subsequently, we fold the first

Brillouin zone (BZ) of the hexagonal lattice, twice, to form a new BZ with 1/3 of its original area, as shown in **Fig. 1(b)**. Consequently, the K (K$^{'}$) point is mapped to the Γ point at the center of the BZ, creating a double Dirac cone. The smaller BZ corresponds to an expanded unit cell in real space of 3 times of the original unit cell area, with 3×2 = 6 masses, and lattice constant $\vec{b_1}$ and $\vec{b_2}$ ($|\vec{b_1}| = |\vec{b_2}| = \sqrt{3}a$), as indicated in **Fig. 1 (a).** The band structure based on the expanded unit cell is plotted in **Fig. 1 (d),** and indicates a double Dirac cone at Γ.

To induce a phase transition, in the topological sense, we break the spatial symmetry of the hexagonal lattice, through changing the spring constants of the connecting masses in the lattice, *i.e.,* distinguishing the *intra* unit cell spring constant $k_1$ from the *inter* unit-cell spring constant: $k_2$. Such distinction still preserves the $C_6$ symmetry of the unit cell. It was found that when $k_1 \neq k_2$, the band degeneracy at the Γ point is lifted and yields a band gap, as indicated in **Fig. 2 (b) and (c)**. With $k_2$ and *m* constant, we continuously change the value of $k_1$ from $k_1>k_2$ to $k_1<k_2$, through which the band gap at Γ point first closes and then reopens. When $k_1=k_2$ , there is no band gap[**Fig. 2 (a) (b)** and **(c)**]. We study the modes related to this transition for (i) $k_1>k_2$ and (ii) $k_1<k_2$.

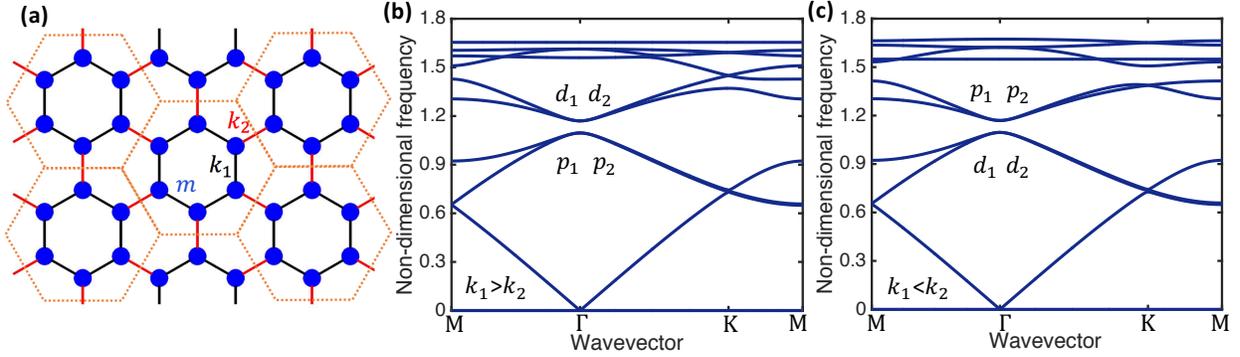

**Figure 2 (a)** Hexagonal spring-mass lattice with intra-cell spring $k_1$ (black straight rods) different from inter-cell spring $k_2$ (red straight rods). **(b)** Band diagram of hexagonal lattice with $k_1>k_2$. Modal displacements at Γ are of $p$ symmetry for the lower degeneracy, and of $d$ symmetry for the higher degeneracy. **(c)** Band diagram of hexagonal lattice with $k_1<k_2$. Modal displacements at Γ are of $d$ symmetry for the lower degeneracy, and of $p$ symmetry for the higher degeneracy.

## III.    RESULTS AND DISCUSSION

### A.  Modal displacement fields in hexagonal spring-mass lattices: The case for pseudospins

The modal displacement and its $x$ and $y$ components, of the masses in the unit cell, at the Γ point of the $k_1>k_2$ lattice are shown in **Fig. 3 (a) – (d)**. The labeling of the modes – in **Figs. 3 (a) – (d)** follows the nomenclature for the lower to higher band degeneracy corresponding to **Figure 2(b)**. The modal displacements for a given mass in $p_1(/d_1)$ are orthogonal to $p_2(/d_2)$, respectively. The constituent $x$ and $y$ direction displacements are plotted successively below. We find that the $x/y$ direction displacements fields at Γ are of odd and even spatial parities – of the $p_x$ (/$p_y$) and $d_{x^2-y^2}$ (/$d_{xy}$) variety, as inferred both from the sense of the displacements and stated relationships in the $C_6$ character table [34]. For instance, the $p_x$ (/$p_y$) character is inferred through reflection in the y- (x-) axis, while the $d_{x^2-y^2}$ (/$d_{xy}$) parity is correlated from the even(/odd) symmetry, with respect to both the $x$ and y axes.

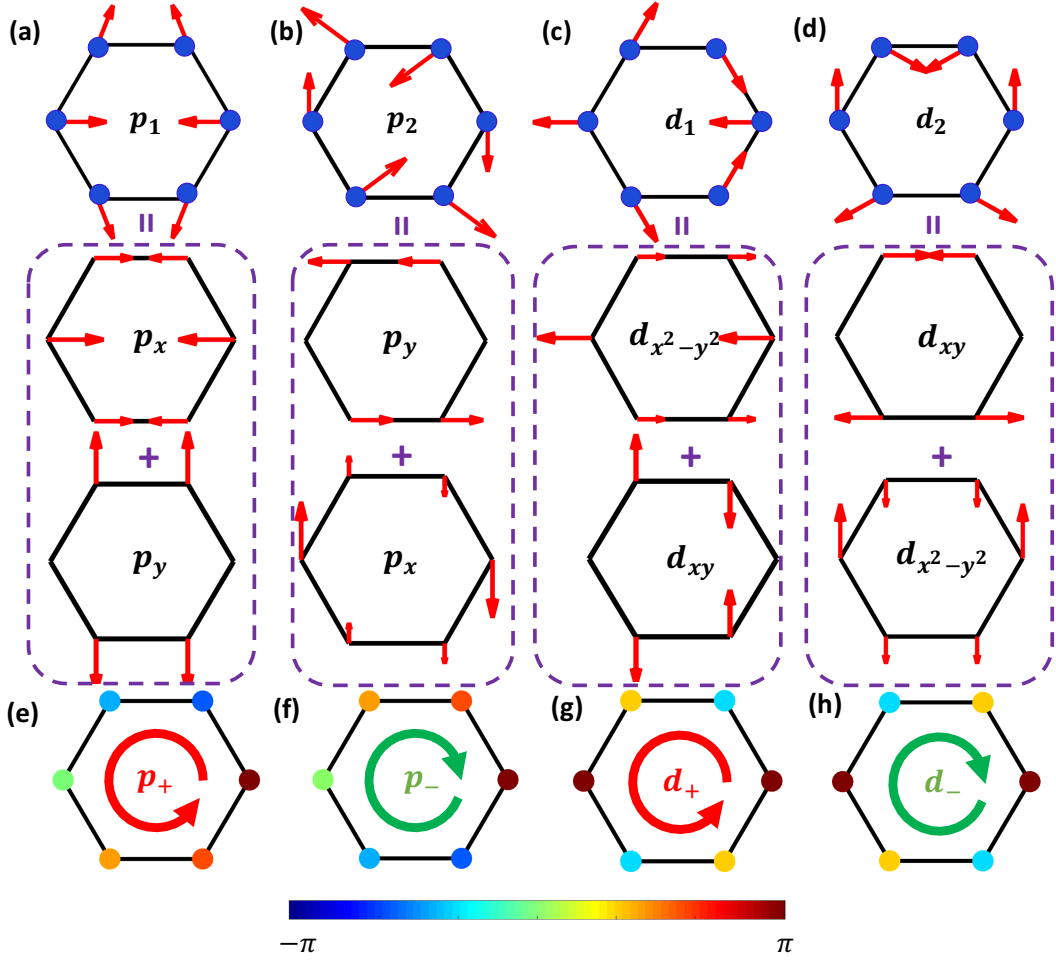

**Figure 3** **(a)** $p_1$, **(b)** $p_2$ and **(c)** $d_1$, **(d)** $d_2$ are total modal displacements for the two two-fold degeneracies at Γ point when $k_1 \neq k_2$. $p_1$ and $p_2$ have odd parities, while $d_1$ and $d_2$ have even parities. $x$ and y direction components to **(a)** and **(b)** clearly show $p_x/p_y$ symmetry, while those to **(c)** and **(d)** that have $d_{x^2-y^2}/d_{xy}$ symmetry. **(e)**, **(f)**, **(g)** and **(h)** are plots of phase relationships between the 6 masses in one unit cell for $p_+$, $p_-$, $d_+$ and $d_-$ in color map, indicating the polarization of wave propagation associated with pseudospin up and pseudospin down.

Hybridizing the $p_1$ and $p_2$ modes in a symmetric and antisymmetric manner yields pseudospins [8],

$$p_\pm = (p_1 \pm ip_2)/\sqrt{2}, \text{ and } d_\pm = (d_1 \pm id_2)/\sqrt{2}. \tag{1}$$

**Fig. 3 (e) – (h)** illustrates the related phase distribution of $p_+$, $p_-$, $d_+$ and $d_-$ in the range of $-\pi$ to $\pi$. Clearly seen from the phase relationship that harmonic wave propagation in $p_+/d_+$ and $p_-/d_-$ have opposite polarizations. Taking the time harmonic component $e^{i\omega t}$ into consideration, due to the orthogonality of displacements in $p_1/d_1$ and $p_2/d_2$, each mass corresponding to the hybridized mode $p_+/d_+$ rotates in the one direction, while each mass in $p_-/d_-$ rotates in the opposite direction. The incorporation of the relative motions of the six masses in the unit cell leads to rotation of the whole displacement field. Such rotation may be considered as one manifestation of a pseudo-spin. One can follow the motion in $d_+$ during one time period $T$: **Figure 4**, indicating such clockwise orientability of the displacement field.

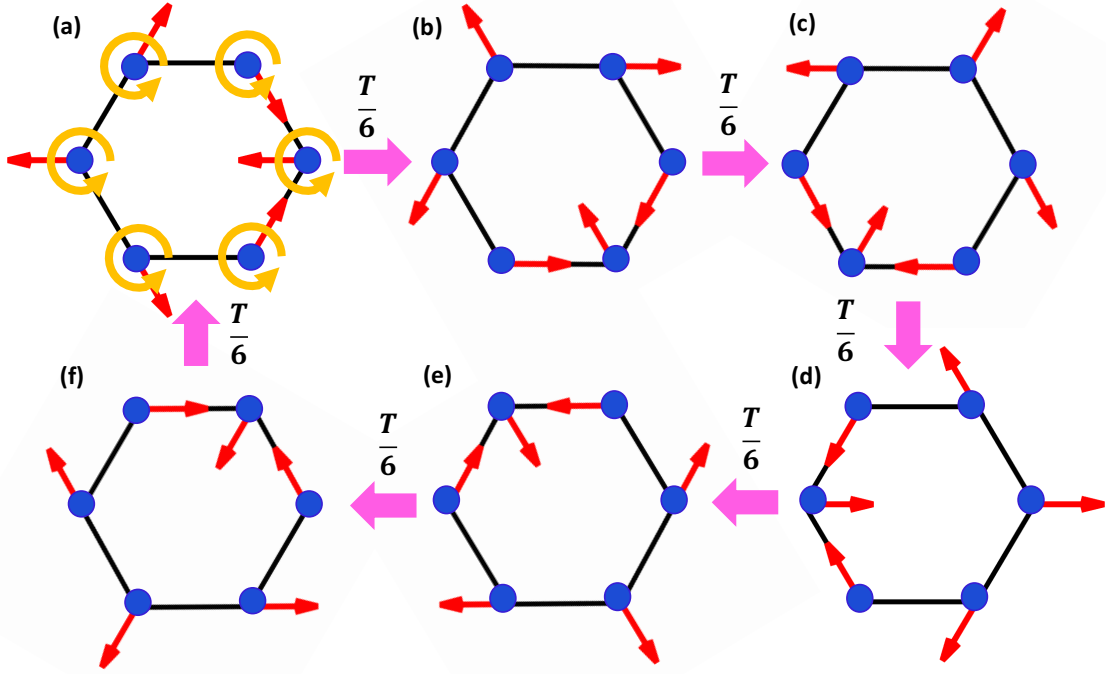

**Figure 4** The spinning of modal displacement field for $d_+ = (d_1 + id_2)/\sqrt{2}$ as a result of time domain motion of the masses during one period $T$.

We find that for the case of $k_1<k_2$, the modal displacement fields have exactly the same odd and even spatial parities, but $d_1$ and $d_2$ are now associated with the higher two degenerate bands, while $p_1$ and $p_2$ corresponds to the lower two bands (**Fig. 2 (c)**). This demonstrates that band inversion happens at the Γ point during the process of closing and reopening the band gap, and a change in topology of the band structure. Such a change has been previously quantified through the spin Chern number [35]. The Hamiltonian on the basis states of $[p_+, d_+, p_-, d_-]$ can be obtained (see Supplementary Material) to be of the following form:

$$H^{eff}(\gamma) = \begin{bmatrix} M - B\gamma^2 & A\gamma_+ & 0 & 0 \\ A^*\gamma_- & -M + B\gamma^2 & 0 & 0 \\ 0 & 0 & M - B\gamma^2 & A\gamma_- \\ 0 & 0 & A^*\gamma_+ & -M + B\gamma^2 \end{bmatrix}, \quad (2)$$

where $\gamma_\pm = \gamma_x \pm i\gamma_y$, and $\gamma^2 = \gamma_x^2 + \gamma_y^2$. $A = i\alpha k_2$ is imaginary ($\alpha > 0$), and $B < 0$. $M = \frac{\varepsilon_d - \varepsilon_p}{2}$ indicates the relative energy of $p$ and $d$ bands, which is positive in the lattice of $k_1 > k_2$, and negative in the lattice of $k_1 < k_2$, respectively. The spin Chern number can be calculated from

$$C_S = \pm \frac{1}{2}\big(\text{sgn}(M) + \text{sgn}(B)\big). \quad (3)$$

Since $B$ is negative, $C_s$ depends on the sign of $M$, which leads to $C_s = 0$ when $M > 0$, and $C_s = \pm 1$ when $M < 0$. This means that for the lattice with $k_1 > k_2$, $C_s = 0$, and the band gap is topologically trivial (**Fig. 2 (b)**). When we decrease $k_1$ to $k_1 < k_2$, the band gap becomes topologically nontrivial (**Fig. 2 (c)**) and $C_s = \pm 1$. Therefore, from the topological band theory [1] it would be expected that there would exist pseudospin-dependent edge modes at the boundary between topologically trivial and topologically nontrivial lattices.

## B. Propagating edge modes

The pseudospin-dependent edge modes are vividly illustrated through simulations on a ribbon-shaped lattice that is periodic in one direction and of the width of one unit cell in the other direction: **Figure 5(a)**. Such a supercell based lattice contains both topologically trivial (T) and nontrivial (NT) units. The NT lattice is constituted from one row of 20 unit cells, and cladded by two T units of 15 unit cells (we chose the number of T and NT units so that the band diagram is relatively scale invariant). Here, the masses in the T and NT units lattice are in the ratio $\frac{m^T}{m^{NT}} = \frac{1.315}{1}$, and spring constants are of the ratio $k_1^T: k_2: k_1^{NT} = 1.2: 1: 0.8$. The inter-cell spring constant $k_2$ is kept the same in both the T and NT units since it connects the two different lattices. The spring constants and masses were chosen such that the T and the NT units have overlapped band gap as related to the frequency ranges indicated in **Fig. 2(b)** and **(c)**. The band structure of the ribbon supper cell is shown in **Fig. 5 (b)** (The frequencies here are non-dimensionalized as $\Omega = \frac{\omega}{\sqrt{\frac{k_2}{m^{NT}}}}$). Compared to the band structures in **Fig. 2 (b)** and **(c)**, we clearly see two additional states appear within the bulk band gap connecting the lower bands to the higher bands, as illustrated by red and green lines in **Fig. 5 (b)**. It was noted that these two new modes propagate with a group velocity of the same magnitude but opposite signs, and correspond to the pseudospin up and pseudospin down topological edge modes. We plotted the modal displacement corresponding to the two additional states of the ribbon lattice near the Γ point ($\gamma = \pm 0.1 \frac{\pi}{b_0}$, $b_0$ is the lattice constant of the extended unit cell) in **Fig. 5 (c)**. These modes are confined to the boundary between the T and the NT units, and decay into the bulk, indicative of edge mode-like character. The appearance of such modes, in the absence of any obvious spin-orbit coupling indicates attributes of a QSHE topological insulator.

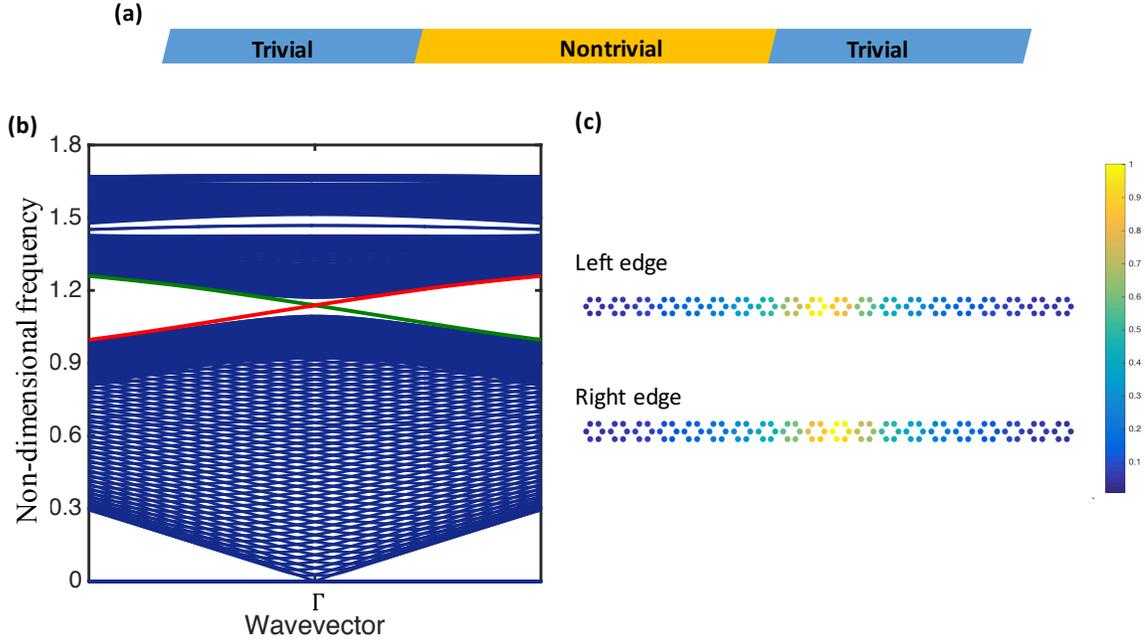

**Figure 5 (a)** Ribbon super cell consists of 20 nontrivial unit cells cladded by 15 trivial unit cells on each end. The mases and springs are of the ratio $\frac{m^T}{m^{NT}} = \frac{1.315}{1}$ and $k_1^T : k_2 : k_1^{NT} = 1.2 : 1 : 0.8$, respectively. **(b)** The band diagram for the ribbon super cell. A pair of pseudospin up and pseudospin down edge modes are found within the bulk band gap (red and green curves). Magnitude of modal displacements of the modes within the bandgap at $\gamma = \pm 0.1 \frac{\pi}{b_0}$ are plotted in **(c)**, from which we can see that they are confined at the edge and decay into the bulk.

To verify the unidirectional propagation of the topological edge modes, we conducted time-domain numerical simulations on finite spring-mass lattices consists of both T and NT units. The governing equation for the spring-mass lattice takes the form $\ddot{u} = Au + F(t)$, where $Au$ is the restoring/displacement-dependent force due to spring deformations, and $F(t)$ is a time-dependent excitation. We solve the equivalent ODE: $\begin{bmatrix} \ddot{u} \\ \dot{u} \end{bmatrix} = AA \begin{bmatrix} \dot{u} \\ u \end{bmatrix} + F(t)$, where $AA = \begin{bmatrix} 0 & A \\ I & 0 \end{bmatrix}$ ($I$ is unitary matrix), using Runge Kutta explicit time integration method (RK4) to determine the displacement $u$ at time $t$. Fixed boundary conditions were applied in the simulations, *i.e.,* masses at the boundaries are connected to springs fixed to the wall.

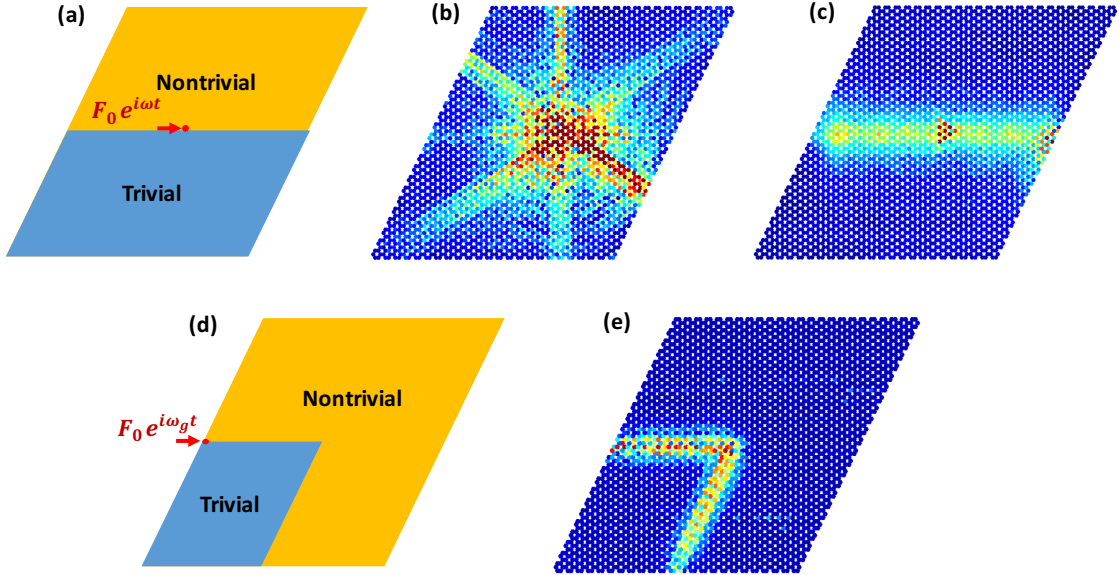

**Figure 6** Time domain simulation of edge wave propagation. The domain is of 24 by 24 unit cells. The mases and springs are of the ratio $\frac{m^T}{m^{NT}} = \frac{1.315}{1}$ and $k_1^T : k_2 : k_1^{NT} = 1.2 : 1 : 0.8$ respectively. **(a)** Sinusoidal excitation force $F = F_0 e^{i\omega t}$ applied on a line edge between topologically nontrivial and trivial spring-mass lattices. **(b)** and **(c)** are the simulation results with $\omega = \omega_b = 0.8\sqrt{\frac{k_2}{m^{NT}}}$ (frequency within the bulk bands), and $\omega = \omega_g = 1.14\sqrt{\frac{k_2}{m^{NT}}}$ (frequency within the bulk band gap). **(d)** is a spring-mass lattice that contains a topological edge with a sharp turning, and **(e)** simulation result on (d) with a force excitation of frequency $\omega_g$.

**Fig 6 (a)** shows the geometry of the modeled spring-mass lattice consisting of a NT and T unit, at the top and bottom, respectively. Initially all the masses are at rest. To avoid boundary reflection, we enforced an excitation force $F(t) = F_0 e^{i\omega t}$ on one of the masses in the NT unit close to the middle of the NT-T boundary, with frequency $\omega = \omega_b = 0.8\sqrt{\frac{k_2}{m^{NT}}}$ corresponding to that of the bulk (from the T/NT band structure), and $\omega = \omega_g = 1.14\sqrt{\frac{k_2}{m^{NT}}}$ corresponding to within the band gap, respectively (for example, a lattice with $m^{NT} = 1$ kg, $k_2 = 10^6$ N/m, $\omega_b = 800$ Hz, and $\omega_g = 1140$ Hz). The simulation results in **Fig. 6 (b)** and **(c)** indicate the amplitude of displacement of the masses, and illustrate that an external force (with $\omega = \omega_b$) will propagate into the bulk, while a force (with $\omega = \omega_g$) will only excite states that propagate at the edge of the T and NT domains. A sharp discontinuity turning boundary between T and NT as indicated in **Fig. 6 (d)** demonstrates that the edge states were immune to backscattering **Fig. 6 (e).**

As the indicated pseudospins are symmetrized configurations of modal displacement fields, they are not prone to selective and individual excitation. However, in another application of the T-NT unit arrangement shown in **Figure 7(a)**, it may be able to separate out the counter-propagating states, as broadly constructed in **Figs. 3 (e) – (h)**. With $F = F_0 e^{i\omega t}$ it was seen that when a left-moving state (say, with positive group velocity) reaches the crossing, it will propagate up to port 1 and down to port 2 along the edges but will not propagate right to the port 3. Consequently, the trajectory of wave propagation (**Figures 7(b) – (f)**) forms a "T" shape. It was noted that the excited modes are sensitive to boundary conditions, that leads to high amplitude at the boundary.

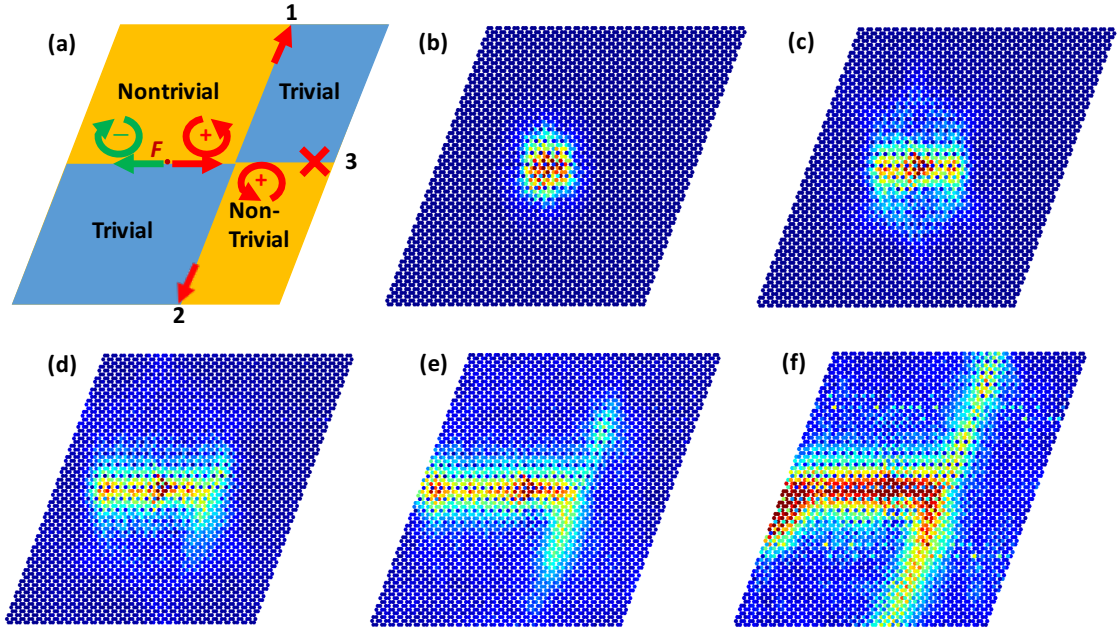

**Figure 7** Pseudospin-dependent wave transport in a waveguide splitter. (a) Waveguide splitter divided into 4 parts, with top left and bottom right of nontrivial lattice, and top right and bottom left of trivial lattice. The cross has angle of 60o to keep the unit cells intact. An excitation force with frequency within the bandgap can excite both pseudospin-up and pseudospin-down modes. Pseudospin-up modes are supported to propagate towards the right, but splits into 2 waves at the cross, that exit from port 1 and 2 respectively. But it is forbidden to exit from port 3, as right-propagate pseudospin-up modes are not supported by the domain to the right of the cross. (b), (c), (d), (e) and (f) are snapshots of time domain simulation at t=$2000t_0$, $4000t_0$, $6000t_0$, $8000t_0$ and $10000t_0$, where $t_0$ is the time step of the simulation.

## IV. CONCLUSIONS

In summary, we have shown that a mass-spring based lattice system may have attributes related to that of a topological insulator, in the presence of time reversal symmetry. Through varying the inter- and inter-unit cell spring constants of such a lattice, for a given mass, a clear and distinct variation of the band structure was seen. A concomitant change in the modal displacement fields, corresponding to a band inversion, may be generated. The deconvolution of the fields as well as their hybridization in a symmetric and antisymmetric manner yields a basis for the creation of pseudo-spins, corresponding to clockwise/counter-clockwise rotation of the modal displacement vector. Both pseudo spin-up and pseudo spin-down modalities, corresponding to the positive or negative group velocity are proposed. The existence of polarized edge states as well as corresponding modes was demonstrated through both frequency domain analysis and time domain simulations. These edge modes are topologically protected, as they are immune to backscattering when encountering sharp edges. Considering that harmonic oscillators (which are direct manifestations of spring-mass units) form the basis for many physical systems, ranging from acoustics to electromagnetics, this work yields a general foundational framework and related methodology, *i.e.,* modulating band structure and constituent modes through varying the respective spring constants of the physical system.

## ACKNOWLEDGEMENTS

This work was supported by ARO grant W911NF-17-1-0453. The authors acknowledge discussions with D. Bisharat and X. Kong.

**References**

[1] M. Z. Hasan and C. L. Kane, Rev. Mod. Phys. **82**, 3045 (2010).

[2] X. L. Qi and S. C. Zhang, Rev. Mod. Phys. **83**, 1057 (2011).

[3] Z. Wang, Y. D. Chong, J. D. Joannopoulos, and M. Soljačić, Phys. Rev. Lett. **100**,013905 (2008).

[4] A. B. Khanikaev, S. Hossein Mousavi, W. K. Tse, M. Kargarian, A. H. MacDonald, and G. Shvets, Nat. Mater. **12**, 233 (2013).

[5] T. Ma, A. B. Khanikaev, S. H. Mousavi, and G. Shvets, Phys. Rev. Lett. **114**, 127401 (2015).

[6] T. Ma and G. Shvets, New J. Phys. **18**, 025012(2016).

[7] L. Lu, J. D. Joannopoulos, and M. Soljačić, Nat. Photonics **8**, 821 (2014).

[8] L. H. Wu and X. Hu, Phys. Rev. Lett. **114**, 223901 (2015).

[9] T. Ma and G. Shvets, Phys. Rev. B **95**, 165102 (2017).

[10] R. Fleury, A. B. Khanikaev, and A. Alù, Nat. Commun. **7**, 11744 (2016).

[11] P. Wang, L. Lu, and K. Bertoldi, Phys. Rev. Lett. **115**, 104302 (2015).

[12] Z. Yang, F. Gao, X. Shi, X. Lin, Z. Gao, Y. Chong, and B. Zhang, Phys. Rev. Lett. **114**, 114301 (2015).

[13] B. Z. Xia, T. T. Liu, G. L. Huang, H. Q. Dai, J. R. Jiao, X. G. Zang, D. J. Yu, S. J. Zheng, and J. Liu, Phys. Rev. B **96**, 094106 (2017).

[14] A. B. Khanikaev, R. Fleury, S. H. Mousavi, and A. Alù, Nat. Commun. **6**, 8260 (2015).

[15] Y. Deng, H. Ge, Y. Tian, M. Lu, and Y. Jing, Phys. Rev. B **96**, 184305 (2017).

[16] C. Brendel, V. Peano, O. Painter, and F. Marquardt, Phys. Rev. B **97**, 020102 (2018).

[17] S. H. Mousavi, A. B. Khanikaev, and Z. Wang, Nat. Commun. **6**, 7586 (2015).

[18] C. He, X. Ni, H. Ge, X. C. Sun, Y. Bin Chen, M. H. Lu, X. P. Liu, and Y. F. Chen, Nat.

Phys. **12**, 1124 (2016).

[19] F. D. M. Haldane, Phys. Rev. Lett. **61**, 2015 (1988).

[20] D. J. Thouless, M. Kohmoto, M. P. Nightingale, and M. Den Nijs, Phys. Rev. Lett. **49**, 405 (1982).

[21] C. L. Kane and E. J. Mele, Phys. Rev. Lett. **95**, 226801 (2005).

[22] B. A. Bernevig and S. C. Zhang, Phys. Rev. Lett. **96**, 106802 (2006).

[23] M. König, L. W. Molenkamp, X. Qi, and S. Zhang, **318**, 766 (2007).

[24] A. Rycerz, J. Tworzydło, and C. W. J. Beenakker, Nat. Phys. **3**, 172 (2007).

[25] D. Xiao, W. Yao, and Q. Niu, Phys. Rev. Lett. **99**, 236809(2007).

[26] Y. Kim, K. Choi, J. Ihm, and H. Jin, Phys. Rev. B **89**, 085429 (2014).

[27] J. Lu, C. Qiu, M. Ke, and Z. Liu, Phys. Rev. Lett. **116**, 093901 (2016).

[28] L. M. Nash, D. Kleckner, A. Read, V. Vitelli, A. M. Turner, and W. T. M. Irvine, Proc. Natl Acad. Sci. U.S.A. **112**, 14495 (2015).

[29] R. K. Pal and M. Ruzzene, New J. Phys. **19**, 025001 (2017).

[30] Y. T. Wang, P. G. Luan, and S. Zhang, New J. Phys. **17**, 073031 (2015).

[31] T. Kariyado and Y. Hatsugai, Sci. Rep. **5**, 18107 (2015).

[32] R. Süsstrunk and S. D. Huber, Science **349**, 47 (2015).

[33] L. Y. Zheng, G. Theocharis, V. Tournat, and V. Gusev, Phys. Rev. B **97**, 060101 (2018).

[34] M. S. Dresselhaus, G. Dresselhaus, and A. Jorio, Group theory: Application to the physics of condensed matter (Springer-Verlag, Berlin, Heidelberg, 2008).

[35] B. A. Bernevig, T. L. Hughes, and S. C. Zhang, Science **314**, 1757 (2006).